\documentclass{vgtc}                          % final (conference style)
\ifpdf%                                % if we use pdflatex
  \pdfoutput=1\relax                   % create PDFs from pdfLaTeX
  \usepackage{graphicx}                % allow us to embed graphics files
  \DeclareGraphicsExtensions{.pdf,.png,.jpg,.jpeg} % for pdflatex we expect .pdf, .png, or .jpg files
\else%                                 % else we use pure latex
  \usepackage{graphicx}                % allow us to embed graphics files
  \DeclareGraphicsExtensions{.eps}     % for pure latex we expect eps files
\fi%

\graphicspath{{figures/}{pictures/}{images/}{./}} % where to search for the images

\usepackage{microtype}                 % use micro-typography (slightly more compact, better to read)
\PassOptionsToPackage{warn}{textcomp}  % to address font issues with \textrightarrow
\usepackage{textcomp}                  % use better special symbols
\usepackage{mathptmx}                  % use matching math font
\usepackage{times}                     % we use Times as the main font
\usepackage{cite}                      % needed to automatically sort the references
\usepackage{tabu}                      % only used for the table example
\usepackage{booktabs}                  % only used for the table example
\onlineid{0}

\vgtccategory{Research}

\vgtcinsertpkg

\title{VegaFusion: Automatic Server-Side Scaling for Interactive Vega Visualizations}

\author{Nicolas Kruchten\thanks{e-mail: nicolas@kruchten.com}\\ %\\ %
\scriptsize École de technologie supérieure
\and Jon Mease\thanks{e-mail: jon@vegafusion.io}\\ % 
 \scriptsize VegaFusion Technologies LLC
\and Dominik Moritz\thanks{e-mail: domoritz@cmu.edu} \\ %
\scriptsize Carnegie Mellon University
}
\teaser{
  \centering
  \includegraphics[width=\linewidth]{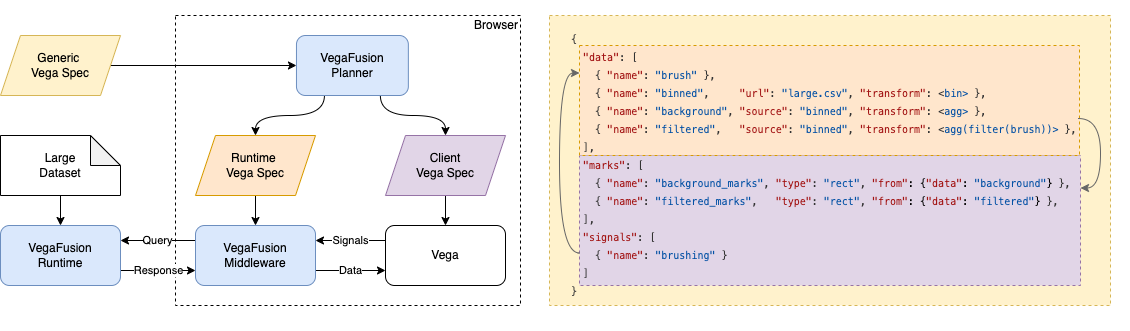}
  \caption{A generic Vega specification is automatically partitioned by the VegaFusion Planner into a runtime specification for the VegaFusion Middleware (describing operations on large datasets) and a client specification for Vega (describing the visualization of the output of these operations as well as client-side interactions). The Middleware dynamically responds to interaction signals from Vega by querying an out-of-browser, natively-compiled VegaFusion Runtime instance and relaying the results back to Vega. The pseudocode on the right illustrates a typical partition for a brushed histogram specification, including the dependencies between data and signals.}
  \label{fig:teaser}
}

\abstract{The Vega grammar has been broadly adopted by a growing ecosystem of browser-based visualization tools. However, the reference Vega renderer does not scale well to large datasets (e.g., millions of rows or hundreds of megabytes) because it requires the entire dataset to be loaded into browser memory. We introduce VegaFusion, which brings automatic server-side scaling to the Vega ecosystem. VegaFusion accepts generic Vega specifications and partitions the required computation between the client and an out-of-browser, natively-compiled server-side process. Large datasets can be processed server-side to avoid loading them into the browser and to take advantage of multi-threading, more powerful server hardware and caching. We demonstrate how VegaFusion can be integrated into the existing Vega ecosystem, and show that VegaFusion greatly outperforms the reference implementation. We demonstrate these benefits with VegaFusion running on the same machine as the client as well as on a remote machine.
} % end of abstract

\CCScatlist{
  \CCScatTwelve{Human-centered computing}{Visu\-al\-iza\-tion}{Visu\-al\-iza\-tion systems and tools}{Visualization toolkits};
  \CCScatTwelve{Human-centered computing}{Visu\-al\-iza\-tion}{Visu\-al\-iza\-tion application domains}{Information visualization};
}

\begin{document}


\firstsection{Introduction}

\maketitle

The Vega~\cite{2014_DeclarativeInteractionDesign} grammar has been broadly adopted as a declarative visualization specification format by a growing ecosystem of tools including Altair~\cite{2018_AltairInteractiveStatistical} for programmatic authoring in Python, Lyra~\cite{2021_LyraDesigningInteractive} for direct-manipulation authoring, and Voyager~\cite{2017_VoyagerAugmentingVisual} for automated recommendation. These tools all use a client-server architecture and integrate the reference implementation of the Vega renderer, which is written in JavaScript and renders visualizations and handles user interactions in a browser. 

The design of the Vega renderer assumes that the entire dataset is available in browser memory, and as a consequence, visualizations based on datasets of more than a million rows or more than 100 megabytes are often unacceptably slow to initially render and exhibit unacceptably high latencies of up to multiple seconds during interaction. Depending on the client hardware, sufficiently large datasets can even cause Vega to crash the containing browser tab.

This problem is particularly vexing when the input data has many rows and the visualization needs to be interactive, but the output does not require displaying one mark per row because it is aggregated. A specific motivating example is the case of cross-filtered histograms for large datasets: each client rendering this visualization using Vega must download and process the entire dataset for initial rendering, and then pass over it again on every brushing interaction. That said, the Vega grammar is declarative, meaning that specification and implementation are not necessarily tightly coupled. This leaves room for the development of new implementations which accept Vega specifications but execute them differently than the reference implementation.

In this paper, we introduce VegaFusion, which brings automatic server-side scaling to the Vega ecosystem. VegaFusion is a new system which works alongside the reference client-side Vega renderer, accepting generic Vega specifications and automatically partitioning the required computation between the client and a server-side process. VegaFusion is open-source software available from \url{https://github.com/vegafusion}.

\section{Background and Related Work}

Vega~\cite{2014_DeclarativeInteractionDesign} is a declarative JSON specification language for interactive data visualizations. The Vega parser parses the specification into a dataflow, which the Vega runtime executes to generate a scenegraph. The scenegraph is then rendered by a Vega renderer into SVG or Web Canvas. To support interactivity, Vega’s dataflow can be parameterized by signals. Vega-Lite~\cite{2017_VegaLiteGrammarInteractive} is a higher-level JSON specification language which compiles down to Vega specifications.

Moritz et al. proposed a comprehensive client-server architecture to dynamically partition a declarative visualization plan based on a cost model and a predictive component to prefetch the results of likely future user interactions~\cite{2015_DynamicClientServerOptimization}, and VegaFusion can be considered the implementation of some of these ideas. VegaPlus~\cite{2022_DemonstrationVegaPlusOptimizing} proposes a similar system to automatically convert parts of the Vega transformation pipelines into SQL statements for server-side processing, albeit with slightly different design goals (notably not R3, see below). Tapestry~\cite{tapestry} uses a similar client-server architecture to scalably embed interactive volume rendering visualizations in web pages.

Falcon~\cite{2019_FalconBalancingInteractive} introduced prefetching and indexing techniques to reduce the latency of cross-filtered histograms, but did not automatically integrate with Vega specifications. SSVG~\cite{ssvg} introduced a lower-level multi-threaded Javascript renderer for SVG. Battle et al. implement machine-learning powered prefetching in ForeCache and demonstrate that it improves performance over naive prefetching~\cite{battle2016dynamic}.

Altair~\cite{2018_AltairInteractiveStatistical} is a Python library which supports the construction of Vega-Lite specifications using a method-chaining API. Notably, Altair provides a mechanism for Python developers to refer to data stored in pandas~\cite{pandas} data frames from a specification and is commonly used within the Jupyter notebook environment~\cite{jupyter} to author visualizations.

\section{Goals and Requirements}

We set out to build a system that could be added to the existing Vega ecosystem with minimal disruption and yet enable Vega-specified visualizations to scale to larger datasets with better initial-render and interaction latencies. We decompose this overarching goal into the following requirements:

\textbf{R1 - Progressive Enhancement, Graceful Degradation, and Parity}: the system must not require authors or tools to construct specifications in a particular way or store their data in a particular system, such as an SQL database. If the system cannot process a given specification, it should gracefully fall back to the Vega reference implementation. Other than performance characteristics, the rendering and interactive behaviour of visualizations should be identical with or without the system.

\textbf{R2 - Moving computation out of the client}: the system should enable computation to be moved out of JavaScript, out of the browser, and ultimately out of the client machine to a physically separate server with more memory, more and faster cores, and/or faster access to the underlying data. That said, the system should also still be able to optionally run entirely on the client machine and ideally still within a browser, albeit with commensurately reduced performance.

\textbf{R3 - Fine-grained caching across interactions}: the system, when running outside the browser, should be able to cache the results of computing individual parts of Vega transformation graphs for reuse in subsequent interactions, even across browser sessions.

\section{Solution}

We designed VegaFusion to meet the above requirements with the three-part architecture outlined in~\autoref{fig:teaser}. This architecture works around and alongside an unmodified Vega implementation (consisting of a parser, runtime, and renderer). The VegaFusion \textbf{Planner} can run in the browser and partitions the incoming generic Vega specification into a client specification which must be run within Vega (e.g., user interaction specifications) and a runtime specification which can run outside of Vega (e.g., aggregations of large datasets). The VegaFusion \textbf{Runtime} can run on the server and handles computations described by the runtime specification, and the VegaFusion \textbf{Middleware} runs in the browser and coordinates communication between Vega and the VegaFusion Runtime. The Middleware is the only component in the VegaFusion architecture which manages the state of the visualization in the browser: the Planner and the Runtime are both stateless, functional processors. The Middleware uses its representation of the state of the visualization to request only the minimal amount of data from the Runtime. All three components are written in Rust, a language which can be compiled to high-performance binary executables to run on servers, or to WebAssembly for execution in a browser.

\subsection{Planner}
The VegaFusion Planner accepts a generic Vega specification and produces two new Vega specifications, one for Vega and one for the VegaFusion Runtime, as well as a communication plan for how Vega events should trigger VegaFusion Runtime calls. Our Runtime currently supports a subset of Vega’s transforms and expression language (see below). Our Planner determines which subset of the computation described in the input specification can be executed by our VegaFusion Runtime and which subset must be executed in Vega. 

For example, as in~\autoref{fig:teaser}, the Vega specification for cross-filtered histograms of columns in a large dataset will include a reference to this large dataset, followed by some transformations such as binning and aggregation calculations, and some signals which bind filter transformations to user-provided brushing operations. In this case, the Planner will determine that all of the operations which are immediately downstream of the large dataset reference, i.e. binning, can be handled by the Runtime, and so the resulting client specification which is passed to Vega does not contain a reference to the full dataset, but rather to multiple smaller intermediate datasets which can be provided by the Runtime: one for the initial rendering and then additional ones for the filtered representations. Datasets which depend on operations not supported by the Runtime, such as stacking, are not extracted into the runtime specification, and will be handled by Vega.

The current implementation of the Planner does not take into account statistics about the data such as its length or cardinality in partitioning the specifications: any operation that can be sent to the Runtime is assigned to the runtime specification. The Planner is the primary way that VegaFusion meets R1: the reference Vega implementation is used for all rendering and interactions, and only the subset of operations that the Runtime supports are executed outside the browser.

\subsection{Middleware}

The VegaFusion Middleware accepts the runtime specification and communication plan produced by the Planner and attaches callbacks to the Vega renderer to subscribe to all signals which serve as inputs to portions of the runtime specification. The Middleware determines which initial queries to the Runtime are required, if any, and thereafter, dynamically calls the Runtime in response to Vega interaction signals to provide all the data required by Vega to render the client specification.

In the case of our cross-filtered histogram example, since the client specification passed to Vega refers to intermediate datasets produced by the Runtime, the Middleware will immediately call the Runtime to produce these intermediate datasets and pass the results to Vega for initial rendering. Thereafter, any time a user interaction such as brushing triggers a Vega signal which triggers a callback as per the communication plan, the Middleware will call the Runtime to produce the new intermediate datasets and pass them back to Vega.

\subsection{Runtime}

The VegaFusion Runtime is the backend server in the VegaFusion architecture: it services queries from the Middleware. VegaFusion queries are computation graph specifications with certain nodes marked for evaluation. The computation graph specification is a directed acyclic graph (DAG) derived by the Middleware from the runtime specification. Root nodes in the DAG (those with no parents) can be data nodes whose values include URLs to data files, or signal nodes whose values include, for example, selection ranges. Non-root nodes describe one of the supported Vega operations such as \texttt{filter}, \texttt{formula}, \texttt{bin}, \texttt{timeunit}, \texttt{aggregate}, \texttt{join-aggregate} and \texttt{window} but do not include their (likely large) values. The role of the Runtime is to compute and return the values for the requested nodes.  An example of a Vega operation not yet supported by the Runtime is \texttt{stack}, which therefore is always run on the client, but it is usually run after aggregation and is not compute-intensive.

\begin{figure}[tb]
 \centering % avoid the use of \begin{center}...\end{center} and use \centering instead (more compact)
 \includegraphics[width=\columnwidth]{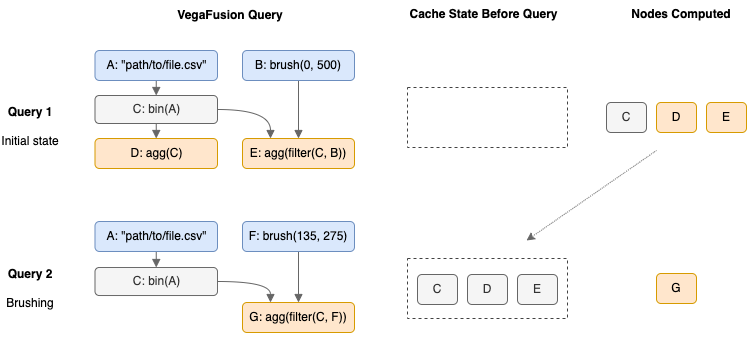}
 \caption{A sequence of two VegaFusion queries along with the state of the cache before the query, and the nodes whose values must be computed to respond to the query. Queries are computation graph specifications with nodes whose values are requested in orange and root nodes in blue. Query 1 contains a data node whose value is the URL of a dataset and whose hashed fingerprint is A, as well as leaf nodes D and E representing the heights of the bars in a histogram, dependent on intermediate node C representing the binned dataset. Node E additionally depends on signal node B, which represents a brush whose initial value is 0-500. When the Runtime receives Query 1, its cache is empty so the values of C, D and E are computed and cached, and the values of D and E are returned as requested. When the user changes the brush range to 135-275, the Middleware will issue Query 2, whose graph is slightly different: the signal node has a new fingerprint F because it has a new value and the requested node has fingerprint G, as it depends on F. Only G's value must be computed because C's value is in the cache. }
 \label{fig:sample}
\end{figure}

Each node has a cheaply-computed fingerprint, which is hashed from a recursive expansion of its definition and that of its parents, all the way to root nodes with values. This fingerprint is computable by both the Middleware and the Runtime, and enables a fine-grained caching scheme. \autoref{fig:sample} shows a schematic view of how caching is used to service a sequence of two VegaFusion Queries. This Runtime caching system is how VegaFusion is able to meet R3. In the example sequence, the large dataset located at the URL which is the value of node A is never transferred to the client; only the smaller, aggregated datasets D, E and F are transferred. Note that the two queries in this sequence need not be issued by the same instance of the Middleware: they could have come from two totally separate client machines, or two browser sessions on the same machine, etc.

The Runtime can run on the same client machine as the browser, or on a remote server so as to meet R2. The Runtime can load datasets from all the same sources as the Vega renderer can, including local and remote CSV and JSON files, which is key to VegaFusion meeting R1.

Datasets are loaded and processed using the Apache Arrow DataFusion~\cite{datafusion} library, whence the name VegaFusion. We selected DataFusion as the core backend technology for VegaFusion because its API was a natural fit for the explicit DAG representation of computation and the caching scheme described above, as well as for its Web Assembly-friendly Rust implementation and its close integration into the popular Apache Arrow ecosystem. Supporting the Vega \texttt{filter} and \texttt{formula} transforms requires supporting Vega's expression language, which is a subset of JavaScript. VegaFusion compiles such expressions into typed DataFusion operations. 

An SQL datastore was an alternative here, but converting VegaFusion queries into SQL would have made it much more challenging to implement the fine-grained caching we sought by forcing us to rely on the SQL datastore's harder-to-control internal caching mechanism: there is no way in SQL to specify which intermediate outputs a query engine should cache. The architecture of VegaPlus~\cite{2022_DemonstrationVegaPlusOptimizing} is quite similar to VegaFusion except for this key design choice, because R3 was not a requirement for VegaPlus.

\section{Demonstration: Scaling Altair}

To demonstrate that the VegaFusion architecture not only meets the decomposed requirements R1, R2, and R3 but can achieve our broader goals of bringing server-side scaling to the Vega ecosystem with minimal disruption, we built a Python library based on VegaFusion which integrates with an unmodified version of Altair. With a single additional line of code, generic examples from the Altair documentation can be scaled to datasets with millions of rows while maintaining smooth interactive performance.

\subsection{Integration}
Altair includes an extension mechanism whereby data transformers can be registered to preprocess data before the Vega-Lite specification is sent to the browser. The VegaFusion extension to Altair registers a data transformer which will write data frames to disk in binary format, automatically launch a Runtime instance and cause it to load data from that file, and rewrite the Vega-Lite specification to refer to this file. The data transformer uses a fingerprint of the data frame to determine filenames to avoid writing the same file to disk more than once. VegaFusion also provides its own browser-side Jupyter rendering extension, which includes the Planner and Middleware components as well as a Vega bundle.

Taken together, the VegaFusion extension to Altair enables a Python developer to load a large dataset into memory in Python using pandas, and then specify, for example, cross-filtered histograms using the Altair API. The Altair API then produces a Vega-Lite JSON specification, which is passed to the VegaFusion renderer via a Jupyter-provided websocket. On the browser, Vega-Lite converts its input into a Vega specification which is intercepted by the Planner and Middleware and Vega is invoked to render the visualization. Any time the user interacts with the visualization, the Middleware calls the Runtime by sending messages back down the Jupyter-provided websocket to the VegaFusion extension Python code which then relays the message to the Runtime process and relays responses back. The large dataset never leaves the machine on which Python is running, and the developer is not responsible for launching or disposing of the VegaFusion Runtime instance. 

The result of this process is that visualizations based on large datasets and specified in Altair not only result in much less data to be transferred to the client and therefore initially render more quickly, but subsequent interactions can exhibit much less latency than with the un-extended version of Altair. 

\subsection{Benchmarks}

We benchmarked Altair versus Altair+VegaFusion in JupyterLab on a 2020 M1 Macbook Air using a figure with three cross-filtered histograms based on a 1 million row dataset stored on disk as a 4.6 megabyte Parquet file. The VegaFusion Runtime and the browser were running on the same machine, so network latency can be assumed to be zero. 

The end-to-end latency from the user executing a notebook cell to initial rendering was 9470ms for Altair and 600ms for Altair+VegaFusion for a speedup in excess of 15. The Altair latency was divided into 5160ms in Python, serializing the dataset then 4300ms in the browser, deserializing and rendering. Altair+VegaFusion spent 250ms in Python writing the dataset to disk and loading it into the Runtime, then 350ms in VegaFusion proper, of which 320ms was spent in the Runtime. Brushing across a histogram in Altair yielded 0.6fps, whereas Altair+VegaFusion rendered at 60fps initially, dropping to 10fps as more data fell within the brush. 

We further benchmarked Altair+VegaFusion on the same dataset and visualization with the Runtime on a physically separate machine using the public Binder~\cite{binder} service. In this test, the client machine was in Eastern Canada, and the server was in Central Europe. Binder kernels run on shared, unspecified hardware, so a direct comparison to the above case is difficult, but comparable Runtime calls were approximately 3 times slower on the server than when running on the client machine. These conditions comprise a worst-case scenario for VegaFusion, with high network latency and a slower server than client.

Initial rendering took 2150ms (530ms on the server, then 1620ms within VegaFusion, of which 1010ms was in the Runtime). The same brushing interaction as above yielded approximately 5fps. This is, as mentioned above, a worst-case scenario for a remote runtime, and locating the server physically closer to the client can significantly reduce latency.

A video demonstrating initial rendering and interaction latencies for the various benchmark cases described above is included as supplementary material.

\section{Discussion} 

VegaFusion is able to better scale computation than the reference Vega renderer in three main ways: by moving computation out of the browser, moving computation to a separate machine, and through caching. Its main limitations are that it cannot help scale all Vega visualizations, it always causes data to be copied into the Runtime, and that its Planner takes into account limited information.

\subsection{Scaling}

VegaFusion moves part of the computation out of the browser's single-threaded JavaScript engine into compiled Rust code which can use multi-threading and specialized batch processor instructions, even on the same machine as the browser. A data scientist running Jupyter on their own workstation can take advantage of VegaFusion's Altair extension to scale visualizations in this mode.

Moving computation out of the browser makes it possible to move it to machines with faster processors, with more cores and memory, and faster access to the underlying data than the client machine. A data scientist using a centrally-hosted, shared installation of JupyterHub, for example, would be using VegaFusion's Altair extension in this mode. Deploying VegaFusion so as to take advantage of a more powerful server machine does come at the cost of network latency on every request to the Runtime as in the benchmarks above. Network latency is typically not so high as to make VegaFusion in this mode slower at initial rendering than Vega alone, since the data set itself does not need to be transferred to the client. During interaction, however, there is a range of smaller dataset sizes where the added network latency can make a VegaFusion visualization less responsive than a Vega one. Neither the Planner nor the Middleware currently take dataset size or network latency into account to counter this problem. Furthermore, VegaFusion does not give a user any more feedback to the user than Vega does that computation is under way, so any unusually long network requests will be perceived as interface lag. 

Although Vega caches the last-computed values of its internal dataflow graph, it does not do so in a long-lived, fingerprinted manner. Consequently, an upstream signal toggling between two values will cause the entire downstream graph to be recomputed on each toggle. The fingerprints used as cache keys by the VegaFusion Runtime depend on the full parent graph of each node, so in this toggling case, each node will only be evaluated once for each value. Furthermore, the Runtime cache can be shared by multiple visualizations, be they in the same browser window or even across different machines, leading to large potential speedups compared to Vega when considering a full user session where each subsequent visualization or interaction can use previously-cached values, potentially including those resulting from the sessions of other users.

\subsection{Limitations}

VegaFusion's biggest limitation is that it cannot help scale visualizations that do not make use of some kind of data reduction such as aggregation or sampling. For example, a Vega specification for a multi-million-point scatterplot will always be run on the client side today. VegaFusion can accelerate a density heatmap of the same data, however, which not only results in better performance but also mitigates the overplotting problem inherent in rendering so many points. That said, VegaFusion today will not automatically convert the specification for such a scatterplot into a heatmap specification. Additionally, although Vega can be used with streaming data, VegaFusion only operates in batch mode on static data.

Second, the Runtime today will always load a copy of the entire dataset into memory, even if the data at rest is stored in a system where it is already indexed and which already supports efficient computation, such as an SQL database or a distributed compute cluster. This provides benefits over loading the data into browser memory as described above, but a truly optimal query plan would explicitly consider the cost tradeoff of copying data vs computing in place.

Finally, the Planner today partitions the computation greedily so as to maximize the amount of work done by the Runtime, without consideration for latency or the size or cardinality of the dataset or the capabilities of the client machine. If the client machine is powerful, the dataset is small, and/or the client-server latencies are high, this can result in lower performance for VegaFusion than by simply using Vega for the entire process.

\section{Future Work}

VegaFusion is a functioning system which can be integrated into various parts of the Vega ecosystem to scale Vega-specified visualizations. There remain many opportunities for further development, however, such as greater coverage of the Vega transform and expression language in the Runtime, and improvements to the Planner so as to have it take into account latency, client capabilities and data statistics in determining the partitioning of the input specification. The Middleware could be extended to include predictive prefetching techniques such as those used in Falcon~\cite{2019_FalconBalancingInteractive}. 

The most promising area is the expansion of the Runtime to directly access datastores which support efficient computation such as SQL databases or Spark clusters and where data might already be stored. Instead of pulling data out of such systems into DataFusion, the Runtime could dynamically compile portions of queries into SQL, as VegaPlus~\cite{2022_DemonstrationVegaPlusOptimizing} does it, or other systems such as Dask~\cite{dask} or Spark~\cite{spark} and let upstream systems handle part of the computation, while the Runtime itself continues to manage the cache. 

Finally, a subset of the Runtime, in particular the cache, could also be compiled to meet the WebAssembly System Interface and run on edge services such as Cloudfare workers. An edge Runtime could perform cache lookups and cheap computations locally to minimize latency and delegate expensive computations to an upstream server Runtime running on more powerful hardware, providing more flexibility in deployment to control latency problems.

\bibliographystyle{abbrv-doi}

\bibliography{paper.bib}
\end{document}